\documentclass[aps,prl,twocolumn,groupedaddress,showpacs,amsmath,
amssymb,amsfonts]{revtex4}
\usepackage{graphicx}
\usepackage{bm}

\begin{document}
\title{Can one count the shape of a drum?}
\author{Sven Gnutzmann$^{3,1}$, Panos D. Karageorge$^2$
and Uzy Smilansky$^{1,2}$}  \email[]{uzy.smilansky@weizmann.ac.il}
\affiliation{$^1$Department of Physics of Complex Systems, The
Weizmann Institute of Science, Rehovot 76100, Israel}
\affiliation{$^2$School of Mathematics, Bristol University,
Bristol BS81TW, England, UK.} \affiliation{$^3$Institut f\"ur
Theoretische Physik, Freie Universit\"at Berlin, Arnimallee 14,
14195 Berlin, Germany}

\date{\today}
\begin{abstract}
Sequences of nodal counts store information on the geometry
(metric) of the domain where the wave equation is considered. To
demonstrate this statement, we consider the eigenfunctions of the
Laplace-Beltrami operator on surfaces of revolution. Arranging the
wave functions by increasing values of the eigenvalues, and
counting the number of their nodal domains, we obtain the nodal
sequence whose properties we study. This sequence is expressed as
a trace formula, which consists of a smooth (Weyl-like) part which
depends on global geometrical parameters, and a fluctuating part
which involves the classical periodic orbits on the torus and
their actions (lengths). The geometrical content of the nodal
sequence is thus explicitly revealed.
\end{abstract}


\maketitle

Eigenfunctions of the Schr\"odinger and other wave equations can
be characterized by the number of their nodal domains - a nodal
domain being a maximally connected region where the eigenfunction
has a constant sign. The intimate connection between the spectra
of wave equations and the corresponding sequences of nodal counts
is well known and frequently used in various branches of physics
and mathematics. Sturm's oscillation theorem states that in one
dimension the $n$-th eigenfunction has exactly $n$ nodal domains.
In higher dimensions Courant proved that the number of nodal
domains $\nu_n$ of the $n$-th eigenfunction cannot exceed $n$
\cite{CH}. Recently, it was shown that the fluctuations in the
nodal sequence $\{\nu_n\}_{n=1}^{\infty}$ display universal
features which distinguish clearly between integrable (separable)
and chaotic systems \cite{BGS}. Their study also leads to
surprising connections with percolation theory \cite{Bogoschmit}.
Moreover, the nodal sequences of several \emph{isospectral} (yet
not isometric) domains were recently shown to differ in a
substantial way \cite{GnutzSS, BandSS}. The later observations
suggest that the nodal sequence stores information about the
domain geometry, and this information is not equivalent to the one
stored in the spectrum. Here, we provide further evidence by
deriving an asymptotic trace formula for the \emph{nodal counting
function}

\begin{equation}
C(K)= \sum_{n=1}^{[\![ K ]\!]} \nu_n \ , \ K>0 \ ;\ \ [\![ \cdot
]\!] = \text{the integer part.}
 \label{eq:nodalcounting_def}
\end{equation}
The trace formula (see
\eqref{eq:torus_cKfluct} and \eqref{eq:nodaltrace} ) shows the explicit dependence of the
nodal sequence on the geometry of the surface in both the smooth
(Weyl-like) and the fluctuating parts. Thus, the nodal trace
formula is similar in structure to the corresponding spectral
trace formula \cite{BerryTabor,cdv, bleher94}. Kac's famous
question ``Can one hear the shape of a drum?'' was triggered by
the study of the progenitors of the \emph{spectral} trace formulae
\cite{kac}. The trace formula for the nodal counts leads us to the
title of this letter in which ``count'' replaces ``hear''. We will
consider here two particular classes of systems, namely, the wave
equation on convex smooth surfaces of revolution and on simple
two-dimensional tori. Generalization to other Riemannian manifolds
in two or more dimensions are possible, provided the wave equation
is separable.

The nodal counting function \eqref{eq:nodalcounting_def} is well
defined if the spectrum is free of degeneracies, $E_n > E_{n-1}$.
In case of degeneracies we represent the wave functions in the
unique (real) basis in which the wave functions appear in product
form.  This, however, does not suffice to set a unique order
within the degenerate states, which consequently introduces
ambiguities in the nodal sequence. To circumvent this problem, we
modify the definition of the nodal counting function: First define
$\tilde{c}(E)=\sum_{n=1}^\infty \nu_n \Theta(E - E_n)$. This
function is based on information obtained from the nodal sequence
and the spectrum. To eliminate the dependence on the latter, we
use the ($\epsilon$-smoothed) spectral counting function
$\mathcal{N}_{\epsilon}(E)= \sum_{n=1}^\infty
(\frac{1}{2}+\frac{1}{\pi} \arctan \frac{E - E_n}{\epsilon})$,
which for finite $\epsilon$ is monotonic and can be inverted. We
define $E_{\epsilon}(K)$ as the solution of
$\mathcal{N}_{\epsilon}(E)=K$ and the modified nodal counting
function is
\begin{equation}
 c(K)=\lim_{\epsilon\rightarrow 0} \tilde{c}( E_{\epsilon} (K) )\ .
 \label{eq:modified_def}
\end{equation}
If there are no degeneracies, $c(K)$ is equivalent to
\eqref{eq:nodalcounting_def} up to a shift $K \rightarrow
K-\frac{1}{2}$. A $g$-times degenerate eigenvalue $E_{n}=E_{n+1}
=\dots=E_{n+g-1}$ contributes  a single step function
$\Theta\big(K-(n-1+\frac{g}{2})\big)\sum_{s=1}^g \nu_{n+s-1}$
where the nodal counting function increases by the sum of the
nodal counts within the degeneracy class. We will derive a trace
formula for this modified nodal counting function below (and omit
the `modified' in the sequel).

We start with the simpler case of a 2-dim torus represented as a
rectangle with side lengths $a$ and $b$ and periodic boundary
conditions. The eigenvalues take the values
$E_{n,m}=(2\pi)^2\left[\frac{n^2}{a^2}+\frac{m^2}{b^2}\right]$
where $m,n\in \mathbb{Z}$. The corresponding wavefunctions for
$m,n \ge 0$ are $\psi_{n,m}(x,y)=\cos(2\pi n x/a)\cos(2\pi m y/b)$
and the cosine is replaced by a sine for negative $m$ or $n$. The
number of nodal domains in the wavefunction $\psi_{n,m}$ is
\begin{equation}
 \nu_{m,n}= (2 |n| +\delta_{n,0})(2|m|+\delta_{m,0})\ .
\end{equation}
The  aspect ratio $\tau=a/b$ is the only free parameter in this
context because the number of nodal domains is invariant to
re-scaling of the lengths.

Using Poisson's summation formula $\sum_{n=n_1}^{n_2} f(n)=
\sum_{N=-\infty}^\infty \int_{n_1-1/2}^{n_2+1/2} {\rm d}
n\ f(n) e^{2\pi
i Nn}$ the leading asymptotic trace formula for the spectral
counting function $\mathcal{N}(E)=\sum_{m,n} \Theta(E-E_{n,m})$
for a torus, is,
\begin{eqnarray}
 \mathcal{N}(E)&=&
 \mathcal{A} E \\
 &+&\sqrt{\frac{8}{\pi}}\mathcal{A}
E^{\frac{1}{4}}\sum_{\textbf{r}}   \frac{
\sin(L_{\textbf{r}}\sqrt{E}-\frac{\pi}{4})}{L_{\textbf{r}}^{\frac{3}{2}}}+
\mathcal{O}(E^{-\frac{3}{4}}) \nonumber
 \label{eq:torus_speccount}
\end{eqnarray}
Here,  $\mathcal{A}=ab/(4\pi)$, and the sum is over the winding
numbers $ \textbf{r}=(N,M)\in\mathbb{Z}^2\backslash(0,0)$ (in the sequel
every sum over $\textbf{r}$ will not include $(0,0)$ unless stated
otherwise).  $L_{\textbf{r}}=\sqrt{(Na)^2+(Mb)^2}$ is the length
of a periodic geodesic (periodic orbit) with $ \textbf{r}=(N,M)$.

Our goal is to derive a similar trace formula for the leading
asymptotic behavior of the nodal counting function. Again, using
Poisson re-summation and the saddle point approximation we get for
 $\tilde{c}(E)$
\begin{eqnarray}
   \tilde{c}(E)&=&
   \frac{2 \mathcal{A}^2}{\pi^2} E^2 \\
   &+&
  E^{\frac{5}{4}} \frac{
  2^{\frac{11}{2}}\mathcal{A}^3}{\pi^{\frac{1}{2}}}
   \sum_{\textbf{r}}  \frac{|MN|}{L_{\textbf{r}}^{\frac{7}{2}}}
\sin(L_{\textbf{r}}\sqrt{E}-\frac{\pi}{4})+ \mathcal{O}(E)\ .
\nonumber \label{eq:torus_cE}
\end{eqnarray}
 To express
the counting function as a function of the index $K$, we formally
invert the spectral counting function to order $\mathcal{O}(K^0)$
\begin{equation}
E(K)=\frac{K}{\mathcal{A}}
-K^{\frac{1}{4}}\frac{2^{\frac{3}{2}}}{\mathcal{A}
\pi^{\frac{1}{2}} }\sum_{\textbf{r}}
\frac{\sin(l_{\textbf{r}}\sqrt{K}-\frac{\pi}{4})}{l_{\textbf{r}}^{\frac{3}{2}}}
\label{eq:invert}
\end{equation}
where $l_{\textbf{r}}=L_{\textbf{r}}/\sqrt{\mathcal{A}}$ is the
re-scaled (dimensionless) length of a periodic orbit. This formal
inversion  needs a proper justification which makes use of the
fact that we actually invert the smooth and monotonic ${\cal
N}_{\epsilon}(E)$. However, a detailed discussion of this point
goes beyond the scope of the present note. The numerical tests
which we provide here, supports the validity of this formal
manipulation. Replacing $E$ by $E(K)$ in \eqref{eq:torus_cE} and
keeping only the leading order terms, we get the nodal counting
function, which we write as a sum
$c(K)=\overline{c}(K)+c_{\mathrm{osc}}(K)$ of a smooth part
$\overline{c}$ and an oscillatory part $c_{\mathrm{osc}}$:
\begin{eqnarray}
\label{eq:torus_cKfluct}
 \overline{c}(K)&=&\frac{2}{\pi^2} K^2 + \mathcal{O}(K)\ ,\nonumber
 \\
c_{\mathrm{osc}}(K)&=&K^{\frac{5}{4}} \sum_{{\textbf{r}}}
    a_{\textbf{r}}
  \sin(l_{{\textbf{r}}}\sqrt{K}-\frac{\pi}{4})
   +\mathcal{O}(K)   \\
a_{\textbf{r}}&=&\frac{2^{\frac{7}{2}}}{\pi^{\frac{5}{2}}
 l_{{\textbf{r}}}^{\frac{3}{2}}} \left(\frac{4\pi^2
|NM|}{l_{{\textbf{r}}}^2}-1\right)\nonumber
\end{eqnarray}
While the smooth part is independent of the geometry of the torus,
the oscillating part depends explicitly on the aspect ratio $\tau$
and can distinguish between different geometries. The main
difficulty in computing higher order corrections to the leading
behavior of the nodal counting function is that products of sums
over periodic orbits appear already in the terms of order $K$.

Turning now to more general surfaces we consider analytic, convex
surfaces of revolution $\mathcal{M}$ which are created by the
rotation of the line $y=f(x), x\in I\equiv [-1,1]$ about the $x$
axis. To get a smooth surface, $f(x)$ in the vicinity of $x= \pm
1$, should behave as $f^2(x) \approx a_{\pm}(1 \mp x)$, with
$a_{\pm}$ positive constants. Convexity is achieved by requiring
the second derivative of $f(x)$ to be strictly negative, and
$f'(x_{max})=0$, $x_{max}\in I$, where $f$ reaches the value
$f_{\mathrm{max}}$. We consider the wave equation $-\Delta
\psi(x,\theta)=E \psi(x,\theta)$ and the Laplace-Beltrami operator
for a surface of revolution is
\begin{equation}
 \Delta = \frac{1}{f(x)\sigma(x)}\frac{ \partial}{ \partial
   x} \frac{f(x)}{\sigma(x)}\frac{ \partial  }{ \partial x}+
 \frac{1}{f(x)^2}\frac{ \partial^2 }{ \partial \theta^2}\ .
 \label{eq:LapBel}
\end{equation}
Here, $\sigma(x)= \sqrt {1+f'(x)^2}$ and $\theta$ is the azimuthal
angle. The domain of $\Delta$ are the doubly differentiable,
$2\pi$ periodic in $\theta$ and non singular functions on $ [I
\times S^1]$. Under these conditions,  $\Delta$ is self adjoint
and its spectrum is discrete. $\Delta $ is separable and the
general solution can be written as a product $\Psi(x,\theta) =
\exp(i m \theta)\ \phi_m(x)$ where $m\in \mathbb{Z}$. For any $m$,
\eqref{eq:LapBel} reduces to an ODE of the Sturm-Liouville type,
with eigenvalues $E_{m,n}$ (doubly degenerate when $m\ne0$) and
eigenfunctions $\phi_{n,m}(x)$ with $n=0,1,2,\dots$ nodes. The
eigenfunctions corresponding to the eigenvalue $E_{n,m}$ can be
written as linear combinations of $\cos(m\theta) \phi_{n,m}(x)$
and $\sin(m\theta) \phi_{n,m}(x)$. To be definite, we chose these
two functions as the basis for the discussion, and associate the
former with positive values of $m$ and the later with the negative
values of $m$. The nodal pattern is that of a checkerboard typical
to separable systems and contains
\begin{equation}
\nu_{n,m} = (n+1)(2|m| + \delta_{m,0})
 \label{eq:nu}
\end{equation}
nodal domains.  The semi-classical spectrum is constructed by
using the Bohr-Sommerfeld approximation \cite{cdv},
\begin{equation}
 E_{n,m}^{\mathrm{scl}} = H(n+\frac{1}{2},m) + \mathcal{O}(1)  \qquad  ,
n \in
 \mathbb{N},\ m\in \mathbb{Z}\ .
 \label{eq:BSspectrum}
\end{equation}
where $H(n,m)$ is the classical Hamiltonian defined in terms of
the action variables $m$ and $n$, where $m$ is the momentum
conjugate to the angle $\theta$ and  $n$ is
\begin{equation}
 \begin{split}
  n(E;m) =& \frac{1}{2\pi} \oint p_x(E,x) \ {\rm d}x =
   \frac{1}{\pi}\int_{x_{-}}^{x_{+}}
   p_x(E,x) {\rm d}x \\
   p_x(E,x)=& \sqrt{(E f^2(x)-m^2)(1+f'(x)^2)}/f(x)\ .
 \end{split}
 \label{eq:naction}
\end{equation}
$x_{\pm}$ are the classical turning points  $E f^2(x_{\pm})-m^2
=0$, with $x_{-}\le x_{max}\le x_{+}$. The hamiltonian is obtained
by expressing $E$ as a function of $n,m$ using the implicit
expression \eqref{eq:naction}. $H(n,m)$ is a homogenous function
of order 2: $H(\lambda n,\lambda m)=\lambda^2 H(n,m) $. It
suffices therefore to study the function $n(m)\equiv n(E=1,m)$
which defines a line $\Gamma$ in the $(n,m)$ plain. $n(m)$ is defined
for $|m| \le f_{\mathrm{max}} $ and is even in $m$,  $n(m)=n(-m)$.
The function $n(m)$ is monotonically decreasing from its maximal
value $n(0)$ to $n(m=f_{\mathrm{max}})=0$. All relevant
information on the geodesics on the surface can be derived from
$n(m)$. Periodic geodesics appear if the angular velocities
$\omega_n=\frac{\partial H(n,m)}{\partial n}$ and
$\omega_m=\frac{\partial H(n,m)}{\partial m}$ are rationally
related. Since $\frac{dn(m)}{dm}=-\frac{\omega_m}{\omega_n}$ this
is equivalent to the condition
\begin{equation}
 M+N \frac{dn(m)}{dm}=0
\label{eq:periodic_motion}
\end{equation}
for $M,N \neq 0$. The integers $\textbf{r} =(M,N)\in
\mathbb{Z}^2\backslash(0,0)$ are the winding numbers in the $\theta$ and
$x$ directions. The classical motion is considerably simplified if
the \emph{twist condition} \cite{bleher94} $n''(m) \equiv
\frac{d^2 n(m)}{dm^2}\neq 0$ for $0<m\le f_{\mathrm{max}}$ is
fulfilled. This excludes, for example, the sphere but includes all mild
deformations of an ellipsoid of revolution. We will assume the
twist condition for the rest of this letter. It guarantees that
there is a unique solution to \eqref{eq:periodic_motion} which we
will call $m_{\textbf{r} }$. Note, that $n'(m)$ has a finite range
$\Omega$ and a solution only exists if $-M/N \in \Omega$. The
cases $N=0$, $M\neq 0$ or $M\neq 0$, $N=0$ are not described by
solutions of \eqref{eq:periodic_motion}. They describe a pure
rotation in the $\theta$ direction at constant
$x=x_{\mathrm{max}}$ where $m_{0,\pm|M|}=\pm f_{\mathrm{max}}$
($N=0$) or a a periodic motion through the two poles at fixed
angle $\theta\ \mathrm{mod}\ \pi$ ($M=0$) such that $m_{|N|,0}=0$.
The  length  of a periodic geodesic is given by
\begin{equation}
L_{\textbf{r} }=2 \pi \left|N n(m_{\textbf{r} }) +  M
m_{\textbf{r} }\right|\ .
\end{equation}
Returning to the spectrum, we note that the leading terms in the
trace formula for the spectral counting function $N(E)=\sum_{m,n}
\Theta(E-E_{n,m})$ can be obtained  by using (\ref{eq:BSspectrum})
and  Poisson's summation formula \cite{bleher94}.
\begin{equation}
 \mathcal{N}(E)= \mathcal{A} E + E^{\frac{1}{4}}
 \sum_{\textbf{r}} \mathcal{N}_{\textbf{r}}(E)
   \label{eq:bleher}
\end{equation}
where
\begin{equation}
\mathcal{A}
 = \int_{-f_{\mathrm{max}}}^{f_{\mathrm{max}}} n(m) {\rm d}m
 =||\mathcal{M}||/4\pi
\end{equation}
and $||\mathcal{M}||$ is the area of the surface. The oscillating
parts contain integrals $\propto
\int_{-f_{\mathrm{max}}}^{f_{\mathrm{max}}} {\rm d}m \ e^{2\pi
iE^{\frac{1}{2}}(N n(m) +M m)}$ which can be calculated to leading
order in $E^{\frac{1}{2}}$ using the stationary phase
approximation. The points of stationary phase are identified as
the classically periodic tori \eqref{eq:periodic_motion} with
$m=m_{\textbf{r}}$. This restricts the range of contributing
$\textbf{r}$ values to the classically accessible domain $-M/N\in
\Omega$.  Thus,
\begin{equation}
\mathcal{N}_{\textbf{r}}(E)=(-1)^{N}
\frac{\sin(L_{\textbf{r}}E^{\frac{1}{2}}+
\sigma   \frac{\pi}{4})}{2\pi |N^3
n''_{\textbf{r}}|^{\frac{1}{2}}}
 +\mathcal{O}(E^{-\frac{1}{2}} )
\end{equation}
where $n''_{\textbf{r}}=n''(m=m_{\textbf{r}})$ and
$\sigma=\mathrm{sign}(n''_{\textbf{r}})$ which is the same for all
values of $\textbf{r}$. The contributions of the terms with either
$N=0$ or $M=0$ or with $-M/N \notin \Omega$  are of higher order
in $1/E$ and will not be considered here.

We are now ready to derive the asymptotic trace formula for
\begin{eqnarray}
   c(K)&=&\tilde{c}(E(K))=\sum_{n=0}^\infty\sum_{m=-\infty}^\infty
  \nu_{mn}\ \theta\left (E(K)-E_{m,n}\right)\nonumber \\
   E(K)&=&\frac{K}{\mathcal{A}}-\left(\frac{K}
     {\mathcal{A}}\right)^{\frac{1}{4}} \sum_{\textbf{r}} \frac{
     \mathcal{N}_{\textbf{r}}(\frac{K}{\mathcal{A}})}
     {\mathcal{A}}+\mathcal{O}(K^0)\ .
\end{eqnarray}
The second equation was obtained by inverting $K=\mathcal{N}(E)$
to the desired order using the trace formula \eqref{eq:bleher}. We
follow the same approach as for $\mathcal{N}$ and expand the
result in $\delta E= E(K)-K/\mathcal{A}$ such that
$c(K)=\tilde{c}(K/\mathcal{A})+\tilde{c}'(K/\mathcal{A})\delta E
+\mathcal{O}(\tilde{c}'' \delta E^2)$ which is consistent if we
neglect all orders smaller than $\mathcal{O}(K)$.  The result can
be expressed as a sum $c(K)=\overline{c}(K)+c_{\mathrm{osc}}(K)$
of a smooth part $\overline{c}$ and an oscillatory part,
$c_{\mathrm{osc}}$, in complete analogy to
(\ref{eq:torus_cKfluct}).
\begin{eqnarray}
 \overline{c}(K)&=&2
 \frac{\overline{mn}}{\mathcal{A}} K^2
 + \frac{\overline{m}}{\mathcal{A}^{\frac{1}{2}}} K^{\frac{3}{2}}
 +\mathcal{O}(K) \nonumber \\
  c_{\mathrm{osc}}(K)&=&
   K^{\frac{5}{4}}
  \!\!\!\!\!\sum_{\textbf{r}:-\frac{M}{N}\in \Omega}\!\!\!\!\!\!
  a_{\textbf{r}}
  \sin(l_{\textbf{r}}K^{\frac{1}{2}}+\frac{\sigma\pi}{4}
   +\mathcal{O}(K)\nonumber\\
  a_{\textbf{r}}&=&(-1)^{N} \frac{m_{\textbf{r}} n(m_{\textbf{r}})
    -2 \overline{mn}}{\mathcal{A}^{\frac{5}{4}}\pi
    |N^3 n''_{\textbf{r}}|^{\frac{1}{2}}}
    \nonumber \\
    l_{\textbf{r}}&=&L_{\textbf{r}}/\sqrt{\mathcal{A}}\\
    \overline{m^pn^q}&=&\frac{1}{\mathcal{A}}\int_{E(m,n)<1}
{\rm d}m {\rm d}n\ |m|^p n^q \nonumber \label{eq:nodaltrace}
\end{eqnarray}
where $l_{\textbf{r}}$ is the re-scaled length of a periodic
geodesic and $a_{\textbf{r}}$ is the amplitude contributed by the
(classically allowed) $\textbf{r}$ torus. For $m_r=0$ or $m_r=\pm
f_{\mathrm{max}}$ only one half of the stationary phase integral
contributes and the amplitude $a_{r}$ has to be multiplied by
$1/2$.

\begin{figure*}
 \begin{center}
   \includegraphics[width=.95\linewidth,height=6.8cm]{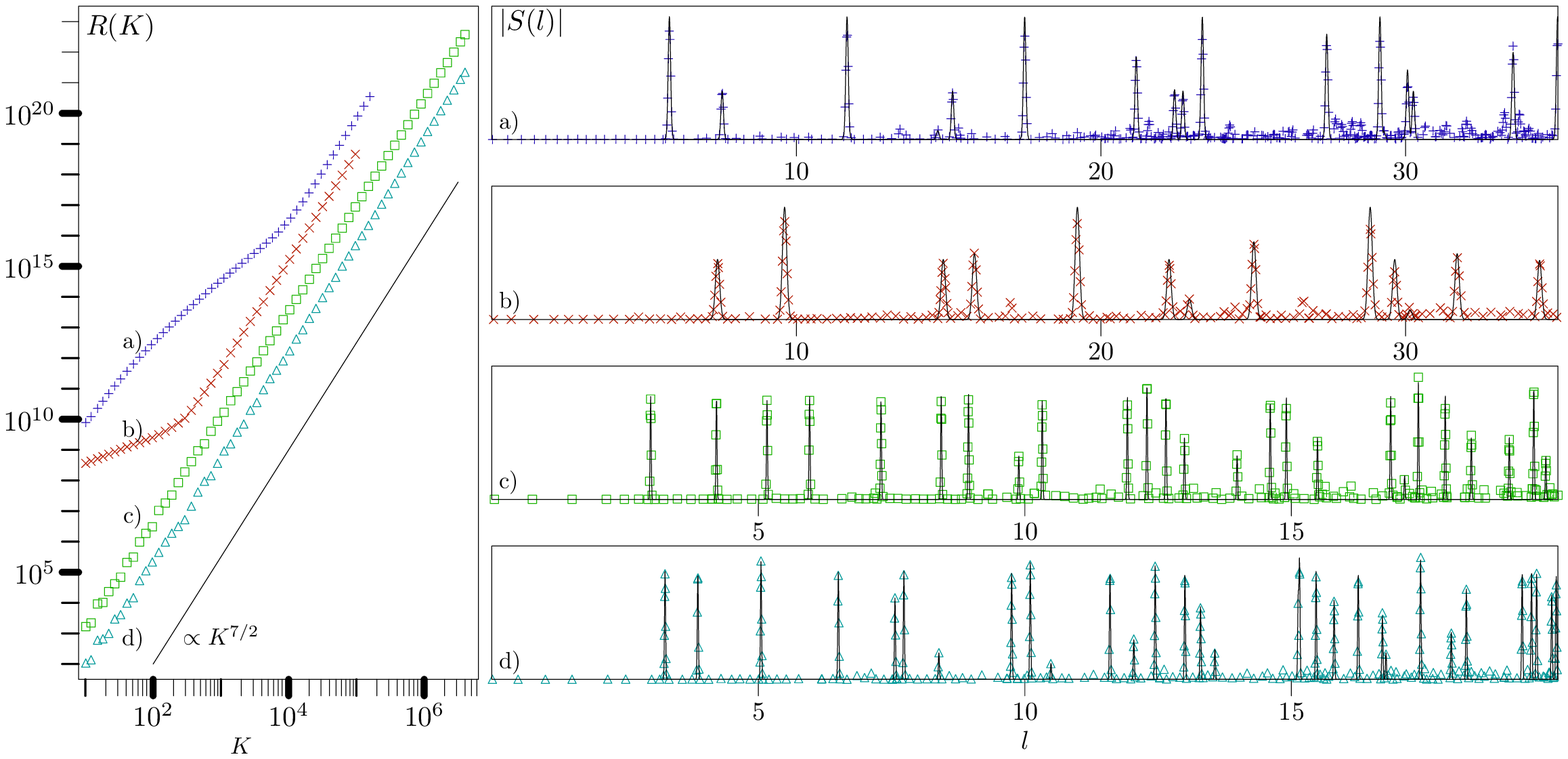}
   \caption{Numerical checks of the  fluctuating parts of trace
     formulae for the two ellipsoids (\ a) and b), see text) and the
     two tori (\ c) and d), see text). \textbf{left}: Double
     logarithmic plot of the integrated squared fluctuations (\ref{rk})
     (arbitrary scale), the full line has slope $7/2$. \textbf{Right}:
     Length spectra of the nodal counting functions (\ref{sl}). The
     full line is obtained from the trace formulae
     \eqref{eq:nodaltrace} and \eqref{eq:torus_cKfluct} and the points
     represent the numerical data.}
 \end{center}
 \label{figure}
\end{figure*}

The approximations involved in the above calculation have been
tested on an extensive numerical data base for two ellipsoids of
revolution defined by the equation $f(x)=R\sqrt{1-x^2}$ ($R=2$ in
data set a), and $R=1/2$ in data set b)) and for two different
tori ($\tau^2=2$ in data set c), and $\tau^2=\sqrt{2}$ in data set
d)). The spectral interval used for the numerical tests included
the first $10^5$ eigenvalues for the ellipsoids, and the first
$4\times10^6$ eigenvalues for the tori. The numerically computed
$c(K)$ were fitted  to a fourth order polynomial in
$\kappa=\sqrt{K}$ and in all cases, the agreement of the two
leading coefficients with the asymptotic theory was better than a
percent. The oscillating part has been obtained numerically by
subtracting the best polynomial fit from the exact $c(K)$. The
fluctuating parts of the trace formulae were tested in two ways.
The integral of the squared oscillatory part
\begin{equation}
 R(K)\approx \int_0^K {\rm d}K'\ c_{\mathrm{osc}}(K')^2\ .
\end{equation}
was computed as a function of $K$, and compared with the
theoretical expression which consists of a double sum over
periodic geodesics. Simplifying this expression by considering
only its diagonal part, one gets the estimate
\begin{equation}
 R(K)=\frac{2}{7}K^{\frac{7}{2}}
 \sum_{\textbf{r}} |a_{\textbf{r}}|^2
 \label{rk}
\end{equation}
which scales like $K^{7/2}$. This scaling has been tested and the
results are shown in the left part of figure \ref{figure}.
Clearly, the expected power law is reached for sufficiently large
values of the counting index $K$. A more stringent test of the
trace formula is provided by computing the length spectrum,
defined by the properly scaled Fourier transform of $c_{osc}(K)$
with respect to $\kappa=\sqrt{K}$.
\begin{equation}
S(l)=l^{3/2}\! \int_0^\infty {\rm d} \kappa\ \kappa^{-5/2}
c_{\mathrm{osc}}(K=\kappa^2)
 e^{-\!\frac{(\kappa-\kappa_0)^2}{\omega}+i \kappa l}
 \label{sl}\ .
\end{equation}
Gaussian windows centered at $\kappa=\kappa_0$ and with a width
$\propto \sqrt{\omega}$ restricted the data used to be well within
the semiclassical domain. The trace formula for the nodal counts
predicts pronounced peaks at the scaled lengths $l=l_{\textbf{r}}$
of the  periodic geodesics. The right frame in Figure \ref{figure}
shows a remarkable agreement of the numerical data with the
theoretical predictions. This excellent agreement provides further
support for the validity of the approximations which were used in
the derivation of the two versions of the nodal counts trace
formula.

Recent studies \cite{GnutzSS,BandSS} have shown that the nodal
sequences of \emph{isospectral} domains are distinct, and can be
used to resolve isospectrality. Thus, the geometrical information
stored in the nodal sequence is not equivalent to the one stored
in the spectral sequence. This result together with the trace
formula obtained here motivates further research of the nodal
sequence as a tool in spectral analysis.

\begin{acknowledgments}
This work was supported by the Minerva Center for non-linear
Physics and the Einstein (Minerva) Center at the Weizmann
Institute, and by grants from the  GIF (grant I-808-228.14/2003),
and EPSRC (grant GR/T06872/01).
\end{acknowledgments}

\end{document}